\documentclass[aps,pra,superscriptaddress,amssymb,amsmath,showpacs,twocolumn]{revtex4}
\usepackage{graphicx}
\begin{document}
\title{Thermalization  of coupled  atom-light  states  in  the
presence  of  optical  collisions}
\author{I.Yu. Chestnov}
\email[]{igor_chestnov@mail.ru}
\author{A.P. Alodjants}
\email[]{alodjants@vlsu.ru}
\author{S.M. Arakelian}
\affiliation{Department of Physics and Applied Mathematics,
Vladimir State University, Gorky str. 87, 600000, Vladimir, Russia}
\author{J. Nipper}
\author{U. Vogl}
\author{F. Vewinger}
\author{M. Weitz}
\affiliation{Institut fur Angewandte Physik der Universit\"at Bonn, Wegelerstra\ss e 8, 53115 Bonn, Germany}


\begin{abstract}
The interaction of a two-level atomic ensemble with a quantized
single-mode electromagnetic field in the presence of optical
collisions is investigated both theoretically and
experimentally. The main focus is on achieving thermal
equilibrium for coupled atom-light states (in particular dressed
states). We propose a model of atomic dressed-state thermalization
that accounts for the evolution of the pseudo-spin Bloch vector
components and characterize the essential role of the spontaneous
emission rate in the thermalization process. Our model shows that
the time of thermalization of the coupled atom-light states
 depends strictly on the ratio of the detuning to the resonant
Rabi frequency. The predicted time of thermalization is in the
nanosecond domain  at full optical power and about 10 times shorter than the natural
lifetime in our experiment.
Experimentally we investigate the interaction of the optical
field with rubidium atoms in an ultrahigh-pressure buffer gas
cell under the conditions of large atom-field detuning comparable
to the thermal energy in frequency units. In particular, an
observed asymmetry of the saturated lineshape is
interpreted as evidence of thermal equilibrium of coupled
atom-light states.
\end{abstract}
\maketitle

\section{INTRODUCTION}
Present remarkable achievements with coherent  manipulation of
coupled  matter-field states  evoke great interest in the
investigation of  phase transitions in  such systems; see, for example,
\cite{1,2,3}. The  key role in the behavior of coupled states
under consideration  is played by   so-called dark and bright
polaritons, that is, bosonic quasiparticles  representing a linear
superposition of  photons in an external (probe) field and the
macroscopic (coherent) polarization  of a two-level atomic system
or excitons localized in quantum wells.

The critical temperature of a phase transition for polaritons can
be high enough due to their  small effective mass, which is   many
orders of magnitude smaller than the free mass of atoms (or
electrons).  Although evidence of  Bose-Einstein condensation
(BEC)  of polaritons in semiconductor microstructures has recently been
reported  by several groups (see \cite{1}, \cite{2}, and \cite{4})
observation of the high-temperature phase transition remains  an
unsolved problem.  In this sense atomic systems seem to be more
attractive and experimentally feasible for polariton BEC purposes \cite{5,6,7,8}.

The main difficulty with polariton condensate observation is
connected with the problem of achieving true  thermal equilibrium
for coupled matter-field states, which is a  primary step
in studying the phase transitions in the systems under consideration.
Roughly speaking, the polaritonic system for the current
experiments with semiconductor microcavities is in nonequilibrium
(or quasiequilibrium) (see, e.g., \cite{9}).  One  mechanism for
achieving thermal equilibrium of low-branch polaritons with the host
lattice is to cool them with a phonon bath \cite{10}.  The thermalization 
time must be shorter than the polariton lifetime in this case. 
For the current experiments the lifetime of
polaritons is in the picosecond regime and comparable to the
thermalization time.

In the area of atomic physics, extremely long lifetimes of
excitations are readily achieved at present. In this paper we 
show that  the atomic polariton lifetime  is limited by the
lifetime of the two-level atomic transition $\tau_{spont}$ only, that is,
by spontaneous emission.  Since the value of $\tau_{spont}$ is in
the nanosecond regime the thermalization time can be longer
compared with that of polaritons formed in semiconductor devices.
Therefore polaritons in an atomic physics system could be
preferable  for observation of BEC, because of
the long coherence times achieved.

Progress toward the  achievement of thermal equilibrium in
coupled atom-field states was made by some of us in
\cite{11}.   In particular, the ability to thermalize coupled
atom-light (dressed) states due to frequent  collisions of
rubidium atoms with buffer gas atoms in the presence of optical
irradiation has been demonstrated experimentally. In the
literature this process is called optical collision (OC) (see,
e.g., \cite{12}). We note that redistribution in OCs 
has also recently allowed laser cooling of
ultradense atomic gases \cite{27}.

In general atomic collisions in the presence of a laser field can
be considered as a scattering  (inelastic) process when both
the internal (or kinetic) energy of the particles and the energy of
the scattered light are changed.  Physically  this  leads to
dephasing of the atomic polarization  that  determines  the
broadening of the fluorescence spectrum  and introduces an
additional (collisional) phase shift. Notably, a closed
representation to understand the effect of atomic collisions on the
resulting width and shape of spectral lines was performed
by Weisskopf in his famous  paper~\cite{13}. Thereafter
collisional broadening  has been  a subject of intensive
investigation both in theory  and in experiment (see,
e.g.,~\cite{14,34,35,36}). Various aspects and approaches to line shape
description, phase shift, and intensity due to nonresonant atomic
collisions have been studied and summarized in \cite{15}. 
Notably, important properties of spectral line shift, width, and asymmetry for a wide range of foreign (buffer) atomic gas densities in connection with the so-called impact limit of atomic collisions have been discussed by A. Royer in \cite{36} (see also \cite{35}).

A statistical approach to the OC problem was proposed
in~\cite{16}, based on the cross section of collisionally aided
radiative excitation (or emission) for two-level atoms coupled to
a thermostat of buffer gas.  Subsequently, in~\cite{17} the authors
established a simple theory of OC based on the dressed-state
approach  in the Schr\"odinger picture.  It has been shown that
collisions with buffer gas particles   reduce  to a transfer
between different dressed states of the atom.   Spectral
redistribution of quasiresonant radiation occurring due to
collisional relaxation has been a subject of  experimental
investigations   as well (see, e.g.,~\cite{18,19,20}).  The OC
spectral line shape is essentially non-Lorentzian, that is,
asymmetric. From a molecular point of view such an asymmetry  can
be understood by taking into account modifications of interaction
potential curves ~\cite{18, 21}.

Although the main features of OCs have  been  investigated for a long
time, the thermodynamic properties of coupled atom-light systems
have not yet been studied fully. It it is pointed out
in~\cite{12} that, in the limit of a low Rabi splitting energy $\hbar
\Omega_R$, that is, for $\hslash\Omega_R\ll k_{B}T$ ($T$ is the
temperature of the two-level atomic ensemble),  the OCs 
reduce to equalizing dressed-state populations
under the secular approximation. The deviations from the Einstein
coefficients for absorption and stimulated emission induced by
OCs are discussed in~\cite{22}. However,  the
problem of thermalization of coupled atom-field states has not been
studied.  Standard theoretical approaches in  this case
are based on the rate equations for the population of dressed states
only, completely ignoring coupling between population and  atomic
coherences  at the same time (see, e.g.,~\cite{11}, \cite{12}, and \cite{22}). Such an
approach seems unsuitable if we keep in mind the problem of observation of
polariton BEC. Actually spontaneous   polarization
buildup  occurs in this case and polariton  coherences become
important (cf.~\cite{9}). Thus,  the role of the secular
approximation in the thermalization process of coupled atom-light
states should be clarified.

The aim of this report is a fundamental theoretical and
experimental investigation of the thermalization of coupled
atom-field states, taking into account spontaneous emission
processes beyond the approximations typically used.

In Sec.\ref{secII} we establish  the model of interaction of
two-level atoms with the quantized   optical field  in the
presence of OCs. Realistic (experimentally
accessible)  conditions for OC of rubidium atoms with  high-pressure 
buffer (argon or helium)  gas  atoms are discussed. In
Sec.\ref{secIII} the  Bloch-like equations for density matrix
elements in the dressed-state basis are derived. Analysis  of
the steady-state solutions of these equations is performed in
Sec.\ref{secIV}. We define the time of thermalization and specify
necessary conditions to achieve thermal equilibrium  for dressed
atom-light states. In Sec.\ref{intensity} we derive the properties
of intensities of spectral components  for an atomic system in the
presence of OC under the atom-field thermalization process. 
Thermalization of coupled atom-light  states  in an ultrahigh-pressure 
buffer gas environment is  experimentally investigated in
Sec.\ref{secVI}. The spectrum of fluorescence obtained from
rubidium atoms is analyzed. We give an explanation for the observed
resonance fluorescence signal  and necessary estimations of the
time of thermalization according to the theoretical approach
developed here. In Sec.\ref{secVII} we summarize our
results and discuss further prospects for the experiment and the
corresponding theory. In particular, the possibility of polariton
condensation is discussed.

\section{\label{secII}MASTER EQUATION APPROACH FOR ATOM-FIELD INTERACTION
      IN THE PRESENCE OF OC}
Atomic collisions in the presence of  a nonresonant radiation field can be considered as the elementary process of a collision between an isolated two-level  atom $A$ and a foreign (buffer) gas atom of sort $B$; that is,
\begin{equation} \label{eq1}
A(a)+B+\hslash\omega_L\leftrightarrows A(b)+B \mbox{,}
\end{equation}
which  occurs with simultaneous  emission (or absorption) of a photon with frequency $\omega_L$. The collision is called an OC when the frequency of the optical  field $\omega_L$ is near resonant  to  an atomic transition in $A$.
\begin{figure}
\includegraphics[scale=1]{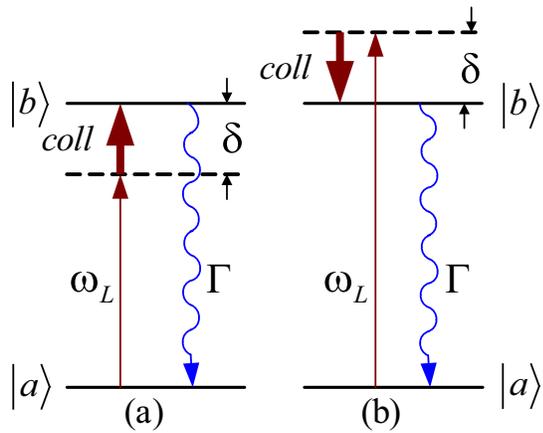}
\caption{\label{fig1}(Color online) Schematic representation of collisionally
aided absorption in a two-level atom  with frequency $\omega_L$
for (a) negative atom-field detuning, $\delta<0$, and (b) 
positive atom-field detuning, that is, $\delta>0$. The decay $\Gamma$ is the
spontaneous emission rate.}
\end{figure}
In Fig.\ref{fig1} schemes of nonresonant absorption of light
with frequency $\omega_L$ during collision with a buffer gas
atom are shown.  The  case of  collisionally aided excitation of 
level $\vert b \rangle$ of the atom for negative atom-light
detuning, $\delta=\omega_L-\omega_0$, is illustrated in
Fig.\ref{fig1}(a). For positive detuning $\delta>0$
($\omega_L>\omega_0$), the atom has some energy excess
$\hbar\delta$, which is transferred to the kinetic  energy of the atoms
after  the collision [Fig.\ref{fig1}(b)]. Note that for both cases
excitation of the upper level ${\left| b \right\rangle} $ is
impossible  due to atomic collision only.

Our description of thermalization  of atom-field (dressed) states
developed here is based on the approach  to OC presented
in~\cite{12} and generalizes their treatment to detuning of
$\delta$ values comparable to or higher than the thermal energy $k_{B}T$.
The master equation for the density matrix $\sigma$ in the
presence of both OCs and radiative (spontaneous)
relaxation  processes can be  written as follows:
\begin{equation} \label{eq2}
\frac{d\sigma }{dt} =-\frac{i}{\hbar } \left[H,\sigma \right]+\left\{\frac{d\sigma }{dt} \right\}_{rad} +\left\{\frac{d\sigma }{dt} \right\}_{coll} ,
\end{equation}
where the last two terms account for spontaneous emission and collisions with buffer gas atoms.
The Hamiltonian  $H$ describes  atom-field interaction under the rotating wave approximation  and has the form
\begin{equation} \label{eq3}
H=\hbar \omega _{L} f^{\dag} f+\hbar \omega _{0} {\left| b \right\rangle} {\left\langle b \right|} +\hbar g\left(S_{+} f+S_{-} f^{\dag } \right),
\end{equation}
where $f (f^{\dag})$  is the annihilation (creation) operator for the photons absorbed (or emitted) due to atomic collisions, $g=\sqrt{{\left|d_{ab}\right|^{2} \omega _{L} }/{2\hbar \epsilon_{0} V}}$  is the atom-field interaction constant, which we take to be identical for all atoms, $d_{ab}$ is the atomic dipole matrix element, and $V$ is the interaction volume.  In expression  (\ref{eq3})  $S_{-} ={\left| a \right\rangle} {\left\langle b \right|} $ and $S_{+} =S_{-}^{\dag } \equiv {\left| b \right\rangle} {\left\langle a \right|} $ represent  atomic  transition operators.
The term
\begin{equation} \label{eq4}
\left\{\frac{d\sigma }{dt} \right\}_{rad} =-\frac{\Gamma }{2} \left(S_{+} S_{-} \sigma +\sigma S_{+} S_{-} \right)+\Gamma S_{-} \sigma S_{+}
\end{equation}
characterizes the contribution of spontaneous  processes  in  Eq.(\ref{eq2}); $\Gamma \equiv {1/\tau _{spont} } $   is the spontaneous emission rate ($\tau _{spont} $ is the natural lifetime of the atomic transition).
The last term in Eq.(\ref{eq2}) describes  the atomic  collisions and can be established as
\begin{equation} \label{eq5}
\left\{\frac{d\sigma }{dt} \right\}_{coll} =-\frac{\gamma }{2} \sigma +2\gamma s_{z} \sigma s_{z} -i\eta \left[s_{z} ,\sigma \right],
\end{equation}
where $s_{z} =\frac{1}{2} \left({\left| b \right\rangle} {\left\langle b \right|} -{\left| a \right\rangle} {\left\langle a \right|} \right)$ is the atomic population inversion operator obeying conditions  $s_{z} {\left| a \right\rangle} =-\frac{1}{2} {\left| a \right\rangle} $ and $s_{z} {\left| b \right\rangle} =\frac{1}{2} {\left| b \right\rangle} $.

In (\ref{eq5}) parameter $\gamma $ characterizes the collisional relaxation rate (collisional broadening) in the presence of a monochromatic laser field; $\eta $  determines the average phase shift appearing due to collisions. In connection with the theory of OC parameters, $\gamma $ and  $\eta $ can be represented as (cf.~\cite{12}):
\begin{subequations} \label{eq6}
\begin{eqnarray}
   \label{eq6a} \gamma =\left\langle 1-\cos \phi \right\rangle _{coll},\\
   \label{eq6b}   \eta =\left\langle \sin \phi \right\rangle _{coll},
\end{eqnarray}
\end{subequations}
\noindent where $\phi =\int _{-\infty }^{+\infty }\left[ \omega _{ba} (t)-\omega _{0} \right]  dt$ is the phase shift that accumulates during the collision.

More rigorous expressions for collisional broadening $\gamma $  can be found using a full quantum mechanical (microscopic) approach (see, e.g.,~\cite{21}).  In general, $\gamma $ implicitly  depends  on the atom-field detuning $\delta $ and, thus, on the molecular potentials for a compound system. For $\delta =0$ the magnitude  of $\gamma $ can be inferred  from   the expression
\begin{equation} \label{eq7}
\gamma \simeq \pi \rho _{0}^{2} {\rm v}_{{\rm T}} N_{B} ,
\end{equation}
where $\rho _{0} $ is  the  Weisskopf  radius, depending on the
molecular level variation (shifting) due to  atomic collisions,
${\rm v}_{{\rm T}} =\sqrt{{2k_{B} T}/{m_{at}}}$  is the
thermal atomic velocity, and  $N_{B}$ is the number density of the
buffer gas.

In particular, in the experiment described in Sec.\ref{secVI}, we
use   rubidium atoms with mass $m_{at} \simeq 1.46\times 10^{-25}
kg$  under the temperature $T=530K$ and a buffer gas of density
$N_{B} \simeq 10^{21} cm^{-3} $. Taking into account typical
values of the Weisskopf radius $\rho _{0} \simeq 10^{-3} \mu m$
for the collisional broadening rate (\ref{eq7}), one can obtain  a
value of few terahertz. The average collisional shift $\eta
$ can be expressed via molecular level variation under  the
quantum mechanical description as well. Practically, $\eta$ in (\ref{eq6b}) 
has the same order of magnitude as a collisional broadening $\gamma $. 
In the current experiment we used argon gas as the buffer
gas (${\gamma/2\pi} \equiv {\gamma _{Ar}/2\pi}
\simeq 7.2~GHz/bar$, ${\eta /2\pi}\equiv {\eta _{Ar}/2\pi} \simeq
- 6~GHz/bar$) at $500~bar$ (50~MPa) pressure and helium as the buffer gas
(${\gamma _{He}/2\pi} \simeq 2.8~GHz/bar$, ${\eta _{He}
/2\pi}\simeq 1.6~GHz/bar$) at $400~ bar$ (40~MPa), respectively.

The spontaneous emission rate for experimentally utilized rubidium
atom D lines is  ${\Gamma} \simeq 2\pi\cdot6$~MHz ($\tau _{spont}
=27$~ns) (see, e.g.,~\cite{29}). Thus,  condition $\gamma
\simeq \eta \gg\Gamma$ is held in the experiment under discussion.
Nevertheless as we will see, spontaneous emission plays
an essential role in  the thermalization process of coupled
atom-light states due to its ability to change the population of
the excited atomic level.

Note that the Doppler effect is not important for the problem
under consideration. First, the motion of the atoms cannot
change the population of atomic levels and, thus, cannot restrict the
thermalization process. Second, Doppler broadening (of the order of
a gigahertz) is significantly  smaller than the collisional broadening in the
experiment, thus neglecting the Doppler broadening in our calculations is justified.

Let us now consider the situation when  each collision happens in
a short enough time span $\tau _{coll} $.  In this case two
collisions are well separated in time; that is, we have
\begin{equation} \label{eq9}
\tau _{coll}\ll T_{coll} ,
\end{equation}
where $T_{coll} $ is the time interval separating two collisions. Here
we neglect temporal correlations between the spontaneous emission
process and the collisional one.  Loosely speaking  spontaneous
emission  and OC optical collision  processes are  well  separated in
time as well. Further, equation (\ref{eq2}) is valid under  the
so-called impact limit of OCs (see, e.g., \cite{12}).
The applicability of  our model in this case is discussed
later.

\section{\label{secIII}BLOCH-LIKE EQUATIONS UNDER THE DRESSED-STATE REPRESENTATION}
In the absence of collisions  with buffer gas atoms,  Hamiltonian (\ref{eq3}) for atom-light interaction explicitly has two eigenstates, called dressed states, defined as
\begin{subequations} \label{eq11}
\begin{eqnarray}
        \label{eq11a} {\left| 1(N) \right\rangle} =\sin \theta {\left| a,N+1 \right\rangle} +\cos \theta {\left| b,N \right\rangle}, \\
      \label{eq11b}   {\left| 2(N) \right\rangle} =\cos \theta {\left| a,N+1 \right\rangle} -\sin \theta {\left| b,N \right\rangle},
\end{eqnarray}
\end{subequations}
where $N$ is the total photon number, and ${\left| a,N+1 \right\rangle} $ and ${\left| b,N \right\rangle} $ are bare atom-light states. The mixing angle $\theta \equiv \theta (\delta )$ is defined by
\begin{equation} \label{eq12}
\tan 2\theta =-\frac{\Omega _{0} }{\delta }, \mbox{\ } 0\leq 2\theta<\pi,
\end{equation}
where   $\Omega _{0} \simeq 2g\sqrt{N} $ denotes the resonant  Rabi frequency.
In general, the parameters $\sin \theta $ and $\cos \theta $  can be  represented  as \cite{11}
\begin{subequations} \label{eq13}
\begin{eqnarray}
        \label{eq13a} \sin \theta =\frac{1}{\sqrt{2} } \sqrt{1+\frac{\delta }{\Omega _{R} } } , \\
      \label{eq13b}  \cos \theta =\frac{1}{\sqrt{2} } \sqrt{1-\frac{\delta }{\Omega _{R} } } ,
\end{eqnarray}
\end{subequations}
where $\Omega _{R} =\sqrt{\delta ^{2} +\Omega _{0}^{2} } $ is the Rabi splitting frequency. It is important  to emphasize  that state ${\left| 2(N) \right\rangle} $ is always located below state  ${\left| 1(N) \right\rangle} $.

\begin{figure}
\includegraphics[scale=0.55]{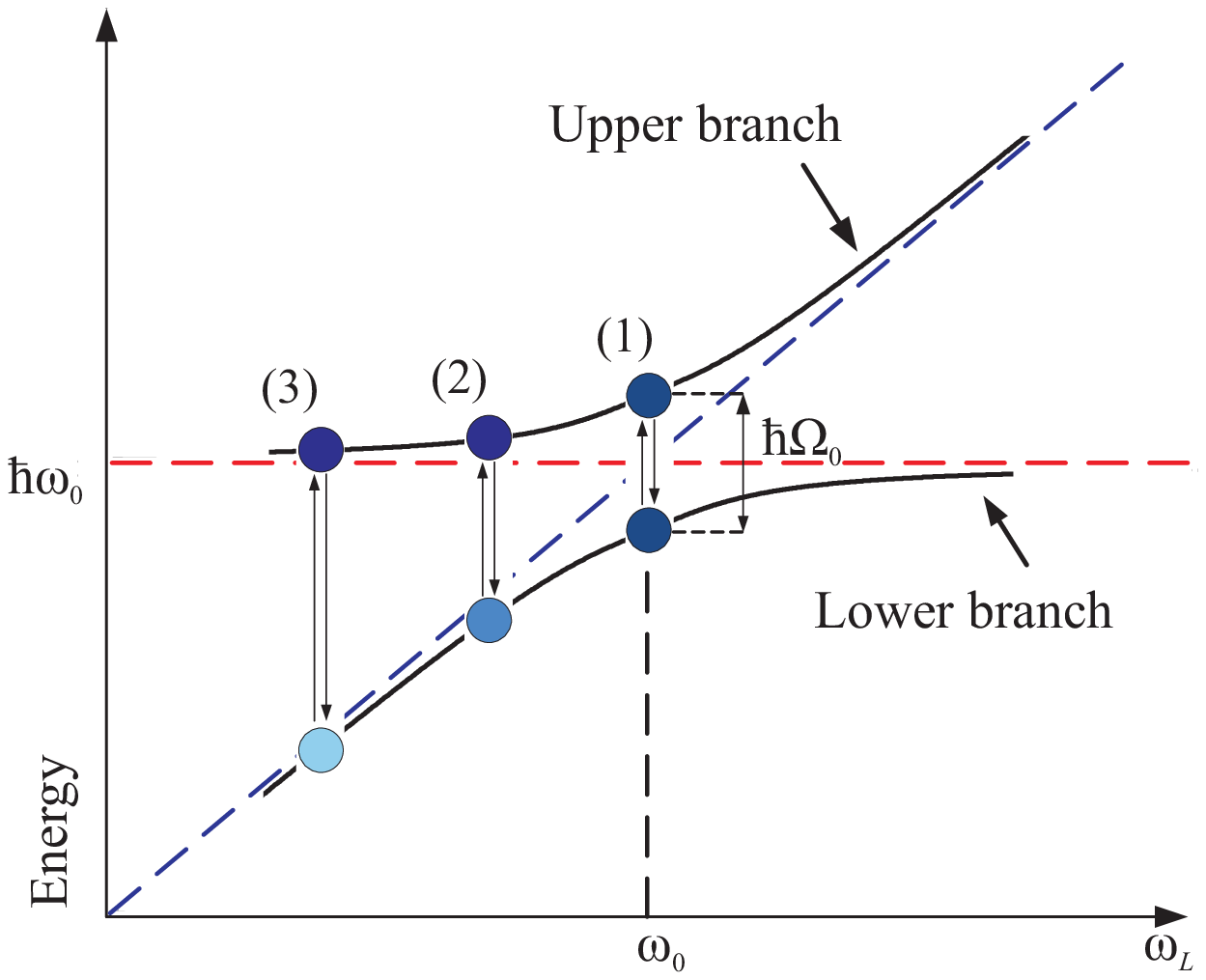}
\caption{\label{fig2} (Color online) Dispersion relation for coupled atom-light
states  versus  $\omega _{L} $; $\hbar \Omega _{0}^{} $ is the
resonant Rabi splitting energy.  The upper branch corresponds to
the ${\left| 1(N) \right\rangle} $ dressed state,  the lower
branch describes the ${\left| 2(N) \right\rangle} $ state. 
The dashed horizontal line corresponds to the
uncoupled level ${\left| b,N \right\rangle} $, that is, to the atomic
transition energy $\hbar \omega _{0}^{} $.  The experimentally
observed coupled states are states (2) and (3),
for which  condition (\ref{eq15}) is satisfied.}
\end{figure}

In Fig.\ref{fig2} the dependence of dressed-atom-state energies as a function of light frequency $\omega _{L} $ is  presented schematically. The frequency gap  between dressed  states ${\left| 1(N) \right\rangle} $ and ${\left| 2(N) \right\rangle} $ is the Rabi splitting  frequency $\Omega _{R}$. The gap is minimal and equal to the Rabi frequency $\Omega _{R,\min } =\Omega _{0}^{} $  under the atom-light resonance condition ($\delta =0$).
In this paper we mostly focus on the case for which the Rabi splitting energy is comparable to or higher than the thermal energy; that is,
\begin{equation} \label{eq14}
\hbar \Omega _{R} \ge k_{B} T.
\end{equation}

For the current experiment  (see Sec.\ref{secVI}) we are limited to Rabi frequencies $\Omega _{0} /2\pi $ by the value of  0.1~THz, which corresponds to full optical power $P_{0} \simeq 300$~mW.  At the same time the thermal energy ($k_{B} T$)  for rubidium atoms  at ambient temperatures ($T=530$~K) corresponds to a frequency of $11$~THz. In this case, we need to have large detuning   $\delta $ values ($|\delta|/2\pi \ge 11$~THz) to fulfill relation (\ref{eq14}). In other words, in this paper we  are practically  interested in the  perturbative limit  when
\begin{equation} \label{eq15}
\Omega_0 \ll |\delta |.
\end{equation}
Physically condition (\ref{eq15}) means that  we deal  with dressed states situated  far from  the region of resonant atom-field  interaction.  In Fig.\ref{fig2}  this situation is  labeled as (2) and (3) respectively.

We solve  the master equation (\ref{eq2}) in the basis of dressed states ${\left| 1(N) \right\rangle} $ and ${\left| 2(N) \right\rangle} $.
The density matrix elements in the  dressed-state representation (\ref{eq11})  traced over the photon number $N$ are defined as
\begin{subequations} \label{eq16}
\begin{eqnarray}
        \label{eq16a} \sigma _{11} =\sum _{N}{\left\langle 1(N) \right|} \sigma  {\left| 1(N) \right\rangle}, \\
      \label{eq16b}  \sigma _{22} =\sum _{N}{\left\langle 2(N) \right|} \sigma  {\left| 2(N) \right\rangle},\\
      \label{eq16d}\sigma _{21} =\sigma _{12}^{*} =\sum _{N}{\left\langle 2(N) \right|} \sigma  {\left| 1(N) \right\rangle}.
\end{eqnarray}
\end{subequations}
The matrix elements $\sigma _{11} $ and $\sigma _{22} $  describe the populations of the dressed states ${\left| 1(N) \right\rangle} $ and ${\left| 2(N) \right\rangle}$, respectively.  The  nondiagonal elements $\sigma _{12} $  ($\sigma _{21} $)  characterize  dressed-state coherences  and correspond to population  transfer  between dressed-state levels. Notably, the total population of dressed states  (\ref{eq16}) is conserved for OC processes; that is, we have ${d\left(\sigma _{11} +\sigma _{22} \right)}/{dt} =0$.

We now consider the properties of the real components $S_{x,y,z} $  of the pseudospin (Bloch) vector $\vec{S}$  combined from matrix elements (\ref{eq16}) as follows:
\begin{subequations} \label{eq17}
\begin{eqnarray}
\label{eq17a} S_{x} =\sigma _{12} +\sigma _{21}, \\
      \label{eq17b}  S_{y} =i\left(\sigma _{12} -\sigma _{21} \right),\\
      \label{eq17c} S_{z} =\sigma _{11} -\sigma _{22}.
\end{eqnarray}
\end{subequations}
With the help of (\ref{eq2})--(\ref{eq5}) it is possible to get Bloch-like equations for pseudospin components $S_{x,y,z}$,
\begin{widetext}
\begin{subequations} \label{eq18}
\begin{eqnarray}
        \label{eq18a} \frac{dS_{x} }{dt} =-\left(\Gamma _{coh} +\varsigma \right)S_{x} -\tilde{\Omega }_{R} S_{y} -2\left(\alpha -\Gamma _{12} \right)\left(S_{z} -\frac{2wS_{z}^{(eq)} }{\left(2w+\Gamma _{+} \right)} \right)  +\Gamma \sin (2\theta ), \\
      \label{eq18b}  \frac{dS_{y} }{dt} =-\left(\frac{\Gamma }{2} +\gamma \right)S_{y} +\tilde{\Omega }_{R} S_{x} +2U\left(S_{z} -\frac{2wS_{z}^{(eq)} }{\left(2w+\Gamma _{+} \right)} \right),\\
      \label{eq18c} \frac{dS_{z} }{dt} =-2w\left[\left(1+\frac{\Gamma _{+} }{2w} \right)S_{z} -S_{z}^{(eq)} \right]-2\left(\alpha -\Gamma _{12} \right)S_{x} -2US_{y} +\Gamma _{-},
\end{eqnarray}
\end{subequations}
\end{widetext}
where the following notations are introduced:
\begin{widetext}
\begin{subequations} \label{eq19}
\begin{eqnarray}
        \label{eq19a} \alpha =\frac{\gamma \sin (4\theta )}{4}, \ \ \ w=\frac{\gamma \sin ^{2} (2\theta )}{2}, \ \ \ \varsigma =\gamma \cos ^{2} (2\theta ), \ \ \  U=\frac{\eta \sin (2\theta )}{2}, \\
      \label{eq19b} \Gamma _{coh} =\frac{\Gamma }{2} \left(1+\sin ^{2} (2\theta )\right), \ \ \Gamma _{\pm } =\Gamma \left(\sin ^{4} (\theta )\pm \cos ^{4} (\theta )\right), \ \ \Gamma _{12} =\frac{\Gamma \sin (4\theta )}{8}.
\end{eqnarray}
\end{subequations}
\end{widetext}

\begin{figure}
\includegraphics[scale=0.9]{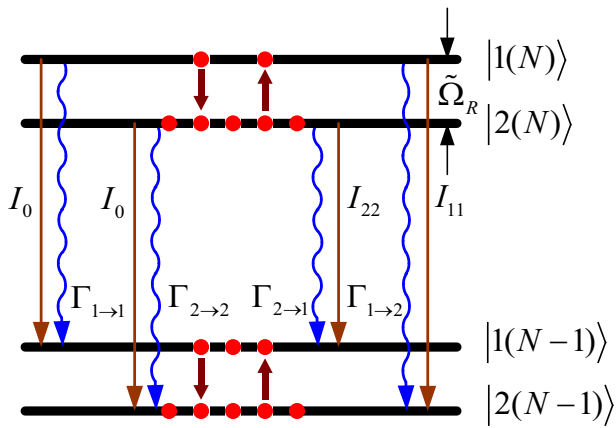}
\caption{\label{fig3} (Color online) Energy (population) transfer between levels of the dressed atom induced by dephasing collisions. $I_{0}$, $I_{11}$ and $I_{22}$ indicate intensity weights of three-color photons emission. In addition, spontaneous emission  described by the rates $\Gamma _{i\to j} $ ($i,j=1,2$)   causes  a transfer between levels of different manifolds.}
\end{figure}

In Eqs.(\ref{eq18}), $\tilde{\Omega }_{R} =\Omega _{R} -{\delta \eta \mathord{\left/ {\vphantom {\delta \eta  \Omega _{R} }} \right. \kern-\nulldelimiterspace} \Omega _{R} } $   is a modified Rabi splitting that takes into account the average phase shift $\eta $ arising due to collisions with buffer gas atoms.  The coefficients $\Gamma _{coh}$, $\Gamma _{\pm } $, and $\Gamma _{12} $ in (\ref{eq19b}) characterize spontaneous emission in the dressed-state representation. In general they  are connected with spontaneous emission rates $\Gamma _{1\to 1} =\Gamma _{2\to 2} =\frac{1}{4} \Gamma \sin ^{2} (2\theta )$, $\Gamma _{2\to 1} =\Gamma \sin ^{4} (\theta )$   and  $\Gamma _{1\to 2} =\Gamma \cos ^{4} (\theta )$  from  dressed-state levels  linking neighbor manifolds (see Fig.\ref{fig3}).
$U$ describes the influence of the average phase shift on the evolution of the system under discussion.

Since we are interested in the conditions for the transition of coupled atom-light states to thermal equilibrium, we have introduced in Eqs.\eqref{eq18}  the  term  ${2wS_{z}^{(eq)} }/{\left(2w+\Gamma _{+} \right)} $  for  the evolution of the Bloch vector component $S_{z}$.  The presence  of such a term in the set of Eqs.\eqref{eq18}  is compatible  with  existing   theories of spin-boson interaction with thermostat particles (due to collisions in our case) discussed by A. Leggett \textit{et al.} in \cite{23}. In particular,  according to our approach in the ideal  case, neglecting spontaneous emission terms  in (\ref{eq18}),  the dressed-state population imbalance  $S_{z}$ should reach its thermodynamically equilibrium value $S_{z}^{(eq)}$   with the rate $2w$ due to collisions with buffer gas atoms  [see (\ref{eq23})], which is in agreement with experimentally tested approaches to OCs  based on the solution of Boltzmann-like (rate) equations \cite{11, 22}.

The dependence of $S_z^{(eq)}$ on temperature can be easily understood from the thermodynamic properties of the coupled atom-light system (cf. \cite{11}, \cite{24}). In particular,  the population of the lower dressed state ${\left| 2(N) \right\rangle} $ should be  much larger [$\exp \left({\hbar \Omega _{R}/k_{B} T} \right)$ times]  than  that of the upper one in thermal equilibrium.  We suppose
\begin{equation} \label{sz_eq}
S_{z}^{(eq)} =-\tanh \left({\hbar \Omega _{R}/2k_{B} T}\right)\approx -\tanh \left({\hbar \left|\delta \right|/ 2k_{B} T} \right),
\end{equation}
where the latter expression is valid under the perturbative limit \eqref{eq15}. In particular, for a near-resonant atom-field interaction (when inequality $\hbar \left|\delta \right|\simeq \hbar \Omega _{R} \ll k_{B} T$ is still true), the equilibrium value of $S_{z}$  according to Eq.(\ref{sz_eq}) is $S_z^{(eq)}=0$ and  we arrive at the well-known result  for which  collisions  with buffer gas atoms  tend to equalize the population of the dressed states (cf.\cite{12}).

In the opposite limit of large detuning one can omit the terms containing collisional broadening $\gamma $ and  spontaneous emission $\Gamma $ from  Eqs.(\ref{eq18}) and 
obtain the same results as in~\cite{24}, obtained from the Schr\"odinger representation for OCs. In particular, without emission the collisions do not induce transitions between  
state ${\left| a,N+1 \right\rangle} $ and state ${\left| b,N \right\rangle} $.  Collisionally aided excitation allows a transfer between dressed-state components also in the case of large detunings. The energy difference is balanced by the kinetic energy of the colliding particles. Since state ${\left| 2(N) \right\rangle} $ is energetically lower than  state ${\left| 1(N) \right\rangle} $, energy is taken from the thermal reservoir of the buffer gas during the transition from ${\left| 2(N) \right\rangle} $ to ${\left| 1(N) \right\rangle} $ (Fig.\ref{fig2}). About $10^{3}$--$10^{4}$ collisions happen during the natural lifetime $\tau _{spont} $ of rubidium atoms.

The population transfer between dressed states  evokes  a
thermalization process of coupled  atom-light states which is
characterized by the thermalization rate  $2w=\gamma \left( {\Omega
_{0}^{2} }/{\Omega _{R}^{2} }\right)  \approx \gamma \left( {\Omega _{0}^{2}
}/{\delta _{}^{2} }\right)  $ in Eq.(\ref{eq18c}) for the dressed-state
population imbalance $S_{z} $.  Notably,  $2w$  depends on  the
collisional rate $\gamma $ as well as on the ratio ${\Omega
_{0}^{}/\delta }$,  which characterizes
the atom-field interaction.  The   dependence of $2w$ on the laser
intensity is in agreement with present theories on OCs 
(see, e.g.,~\cite{21}), whose validity in the yet
unexplored high-pressure buffer gas regime with a high
multiparticle collisional rate, however, remains to be tested.

\section{\label{secIV}THERMALIZATION OF COUPLED ATOM-LIGHT  STATES}
Our goal in this section  is to find stationary solutions of Eqs.(\ref{eq18})  which are close to the thermodynamically true equilibrium state $S_{z}^{(eq)} $ of the coupled atom-field system.

The full set of Eqs.\eqref{eq18} yields steady-state solutions:
\begin{widetext}
\begin{subequations} \label{st_solve}
\begin{eqnarray}
S_{x}^{(st)} =-\frac{1}{D} \left[-\Gamma \sin 2\theta \left(\frac{\Gamma }{2} +\gamma \right)+2(U\tilde{\Omega }_{R} +(\alpha -\Gamma _{12} )(\frac{\Gamma }{2} +\gamma ))\left(S_{z}^{(st)} -\frac{2wS_{z}^{(eq)} }{2w+\Gamma _{+} } \right)\right], \\
S_{y}^{(st)} =\frac{1}{D} \left[2(U(\Gamma _{coh} +\varsigma )-\tilde{\Omega }_{R} (\alpha -\Gamma _{12} ))\left(S_{z}^{(st)} -\frac{2wS_{z}^{(eq)} }{2w+\Gamma _{+} } \right)+\tilde{\Omega }_{R} \Gamma \sin 2\theta \right], \\
S_{z}^{(st)} =\frac{2wS_{z}^{(eq)} }{2w+\Gamma _{+} } +\frac{D\Gamma _{-} -2\Gamma \sin 2\theta [(\alpha -\Gamma _{12} )(\frac{\Gamma }{2} +\gamma )+U\tilde{\Omega }_{R} ]}{(2w+\Gamma _{+} )D-4(2U\tilde{\Omega }_{R} (\alpha -\Gamma _{12} )+(\alpha -\Gamma _{12} )^{2} (\frac{\Gamma }{2} +\gamma )-U^{2} (\Gamma _{coh} +\varsigma ))}, 
\end{eqnarray}
\end{subequations}
\end{widetext}
where we made the denotation $D\equiv {\tilde{\Omega }_{R}}^2+(\Gamma_{coh}+\varsigma)\left[ ({\Gamma}/{2})+\gamma\right] $.

First, we examine the role of atomic collisions in the thermalization process,  completely neglecting spontaneous emission within the so-called secular approximation:
\begin{equation} \label{eq22}
\Gamma \ll {\rm \; }\gamma ,{\rm \; }\eta \ll \Omega _{R}.
\end{equation}
In this case from expressions \eqref{st_solve} one can assume that
\begin{subequations} \label{eq23}
\begin{eqnarray}
        \label{eq23a} S_{x}^{(st)} \approx S_{y}^{(st)} \approx 0, \\
      \label{eq23b}  S_{z}^{(st)} \approx S_{z}^{(eq)}.
\end{eqnarray}
\end{subequations}
The inequalities \eqref{eq22} represent the necessary condition for achieving a thermodynamically true equilibrium (\ref{eq23}b) for the dressed-states population. The coupling between dressed-state coherences $\sigma _{12} $ ($\sigma _{21} $) and populations  $\sigma _{11} $, $\sigma _{22} $ can be completely neglected in this case; see (\ref{eq23}a) (cf.~\cite{11}).

\begin{figure}
\includegraphics[scale=0.6]{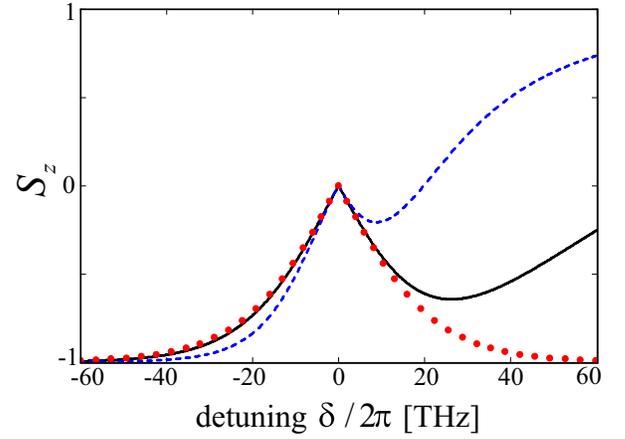}
\caption{\label{fig4} (Color online) Dependence of dressed-state population
imbalance $S_{z} $  as a function of detuning $\delta /{2\pi} $
for  500-bar argon gas collisional broadening $\gamma
/{2\pi}=$~3.6~THz; collisional shift $\eta/2\pi=-3$~THz. The resonant
Rabi splitting frequency $\Omega _{0}/{2\pi} $ is infinite for the
dotted (red) curve,  0.1~THz  for the solid  (black) curve, and
 0.03~THz  for the dashed (blue) curve. In all cases the gas  temperature
is  $T=$~530~K and the spontaneous emission rate is $\Gamma \simeq
37$~MHz.}
\end{figure}

The thermodynamically  full  equilibrium  behavior of the dressed-state population imbalance $S_{z}^{(eq)} $ is represented by the dotted (red) curve in Fig.\ref{fig4}.  It is important to emphasize that  the atomic ensemble under discussion  indicates a two-level system without inversion under the perturbative limit  (\ref{eq15})  for negative atom-field detuning $\delta $ only. Actually in this limit from dressed-state definitions \eqref{eq11}, we have $\sin \theta \simeq 0$,  $\cos \theta \simeq 1$,  and  the lower dressed state   ${\left| 2(N) \right\rangle}$, which is macroscopically occupied,   corresponds to the bare state ${\left| a,N+1 \right\rangle}$, which describes atoms in the ground state.

For large positive atom-light field detuning  $\delta >0$ under condition (\ref{eq15}), we can put $\sin \theta \simeq 1$,  $\cos \theta \simeq 0$, in Eqs.\eqref{eq11}, which implies that  the lower dressed state  ${\left| 2(N) \right\rangle} $ corresponds to the excited atomic level ${\left| b,N \right\rangle}$, which is much more populated. In this case we achieve  inversion in the two-level  atomic system being under thermal equilibrium  (cf.~\cite{24}).

The decay of atomic  upper level ${\left| b \right\rangle} $ with the rate $\Gamma $ leads to nonequilibrium processes in the dressed-state population behavior. Taking into account (\ref{eq18}c) spontaneous emission  under the assumption (\ref{eq23}a), we get a steady-state solution for the dressed-state population imbalance $S_{z} $:
\begin{equation} \label{eq24}
      S_{z} =\frac{2w(\theta )S_{z}^{(eq)} +\Gamma _{-}(\theta )}{2w(\theta )+\Gamma _{+}(\theta )}.
\end{equation}

In Fig.\ref{fig4}  the dependence of the dressed-state population imbalance $S_{z}$ in the steady state (\ref{eq24}) on the atom-field detuning  $\delta $ is plotted.  The full thermalization   of the atom-light state  (dotted curve in Fig.\ref{fig4}) is approached at  infinite Rabi frequency (i.e. for infinite input  laser power) only.   The same result  can be obtained from Boltzmann-like (rate) equations for dressed-state populations  if the contribution from spontaneous emission  to the thermalization of the atom-field states is completely neglected~(cf.~\cite{11}).
The asymptotic behavior of population imbalance presented in Fig.\ref{fig4}  for large values of atom-field detuning $\delta $  can be easily  analyzed in the perturbative limit (\ref{eq15}) under condition (\ref{eq14}). In particular, from (\ref{eq24}) one  can obtain
\begin{subequations} \label{eq25_1}
\begin{eqnarray}
S_{z} \approx S_{z}^{(eq)} +\frac{\frac{\Omega _{0}^{4} }{8\delta ^{4} } -2e^{-{\hbar \left|\delta \right|/ k_{B} T}  }}{1+\frac{\gamma \Omega _{0}^{2} }{\Gamma \delta ^{2} } } \mbox{  for } \delta < 0, \\
S_{z} \approx S_{z}^{(eq)} +\frac{2}{1+\frac{\gamma \Omega _{0}^{2} }{\Gamma \delta ^{2}}} \mbox{ for } \delta > 0.
\end{eqnarray}
\end{subequations}

The   terms containing  $\Gamma $ in  (\ref{eq24}) and (\ref{eq25_1})   characterize the influence of spontaneous emission on the thermalization process. The decay of the dressed-state population imbalance in (\ref{eq24})  due to spontaneous emission is described by $\Gamma _{+}$. To minimize this effect we require the rate of  thermalization of atom-field dressed state $2w$ to be much higher than the effective rate of spontaneous emission $\Gamma _{+} $; that is, the inequality
\begin{equation} \label{eq25}
2w \gg \Gamma _{+}
\end{equation}
should be satisfied.

With the help of definitions (\ref{eq19}) and (\ref{eq25}), one can obtain
\begin{equation} \label{eq26}
\frac{\Gamma }{\gamma } \ll \frac{\Omega _{0}^{2} }{\delta ^{2} }\ll 1.
\end{equation}
Relation (\ref{eq26})  is one of the main results of the paper and a characterizing condition  for reaching true  thermal equilibrium of coupled atom-field (dressed) states.  Such an equilibration is achieved  at the far detuned tails of  dressed-states population imbalance  $S_{z}^{} $ in Fig.\ref{fig4}, for which, with condition (\ref{eq25_1}), we have $S_{z} \approx S_{z}^{(eq)} $.

The suppression of the thermalization process due to spontaneous emission leads to the formation of thermodynamically quasiequilibrium of coupled atom-light states for which condition \eqref{eq26} is violated. Indeed, for
\begin{equation} \label{eq27}
\frac{\Omega _{0}^{2} }{\delta^{2} } \ll\frac{\Gamma }{\gamma } \ll 1
\end{equation}
from \eqref{eq25_1} we get
\begin{subequations} \label{eq27_1}
\begin{eqnarray}
S_{z} \approx S_{z}^{(eq)} +\frac{\Omega _{0}^{4} }{8\delta ^{4} } -2e^{{-\hbar \left|\delta \right|/ k_{B} T} } \mbox{\ for }\delta <0, \\
S_{z} \approx S_{z}^{(eq)} +2 \mbox{\ for }\delta >0.
\end{eqnarray}
\end{subequations}

For highly nonequilibrium coupled atom-light states  $S_{z} \approx 1$ [see (\ref{eq27_1}b) and Fig.\ref{fig4}]  and  spontaneous emission decay drives the dressed-state system out of equilibrium for large positive detuning $\delta $.
In conclusion, Eqs.\eqref{eq18} yield a stationary solution only.

Conditions (\ref{eq25}) and (\ref{eq26}) can be represented in some other form introducing the time of thermalization $T_{therm} $ of coupled atom-field states as follows:
\begin{equation} \label{eq28}
T_{therm} \equiv \frac{2\pi }{2w} =\frac{2\pi }{\gamma } \left(1+\frac{\delta ^{2} }{\Omega _{0}^{2} } \right).
\end{equation}
According to this definition the minimal time of thermalization $T_{therm}^{(\min )} \simeq {2\pi /\gamma } $  occurs  for a near-resonant atom-field interaction  ($\left|\delta \right|\ll\Omega _{0}^{} $) representing state (1) in Fig.~2.  The time $T_{therm}^{(\min)} $ is  in the picosecond regime for experimentally accessible collisional broadening $\gamma$.

For large atom-light detuning in the perturbative limit (\ref{eq15}), the  time of thermalization $T_{therm} $  increases and approaches 
\begin{equation} \label{eq29}
T_{therm} \approx \frac{2\pi \delta ^{2} }{\gamma \Omega _{0}^{2} } \simeq \frac{2\pi k_{B}^{2} T^{2} }{\gamma \hbar ^{2} \Omega _{0}^{2} }.
\end{equation}

Notably, the quantity $T_{therm} $ is inversely proportional to $\Omega _{0}^{2} $, that is, to  the incident  optical power $P_{0} $.
Taking into account definition (\ref{eq28}), we can express condition (\ref{eq25}) for  thermalization of atom-field coupled states as
\begin{equation} \label{eq30}
T_{therm} \ll\tau_{spont}.
\end{equation}
Thereby, at  appropriate time scales the  thermalization time $T_{therm} $  of the atom-field dressed states must be faster than the atomic transition  lifetime $\tau _{spont}$.  In particular, for our experiment the minimal  achievable  time of thermalization at full optical power is about 3.37 ns for argon buffer gas and 10.8 ns for helium (see Sec.\ref{secVI}).

Now let us examine nonequilibrium properties of the atom-light states.  The value of detuning $\delta $ is of the order of the collisional broadening $\gamma $ in our experiment. In fact, we are working beyond the secular approximation limit established by (\ref{eq22}). In this case the Bloch components $S_{x} $ and $S_{y} $ can be  adiabatically eliminated from  Eqs.(\ref{eq18}) and we can obtain the following solution for the dressed-state population imbalance  $S_{z}$:
\begin{equation} \label{eq31}
S_{z} (t)=S_{z}^{(st)} +\left(S_{z} (0)-S_{z}^{(st)} \right)e^{-(2w+\Gamma _{+} )t}  ,
\end{equation}
where $S_{z} (0)$ is the initial value of the pseudospin component $S_{z}$ at $t=0$, and $S_{z}^{(st)} $ is determined by Eq.(\ref{st_solve}c).

\begin{figure}
\includegraphics[scale=0.8]{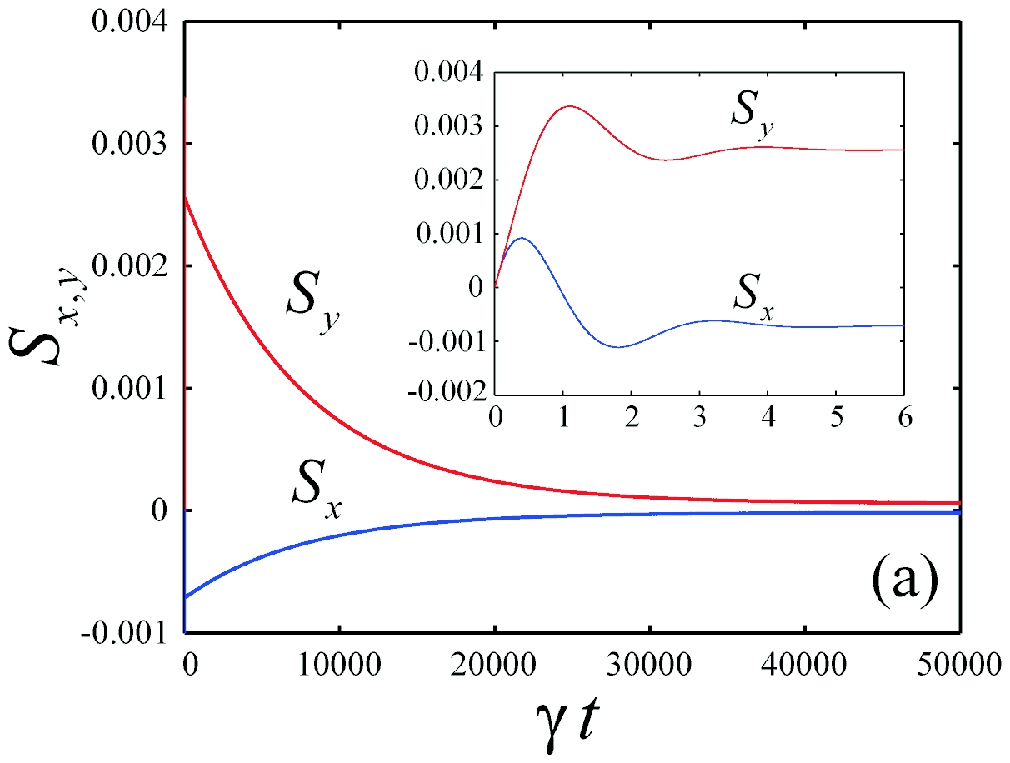}
\includegraphics[scale=0.8]{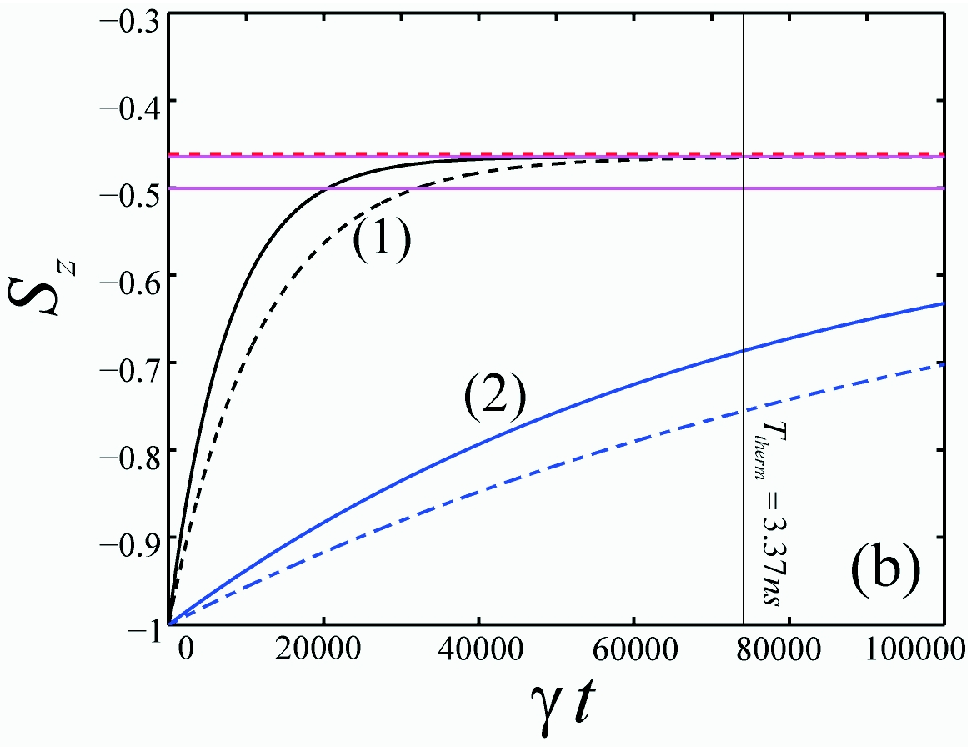}
\caption{\label{fig5} (Color online) Dependences of  (a) dressed-state pseudospin components $S_{x,y} $ and (b) population imbalance  $S_{z} $  as a function on reduced time $\gamma t$. Parameters are: $\gamma/2\pi=3.6$ THz, $\eta/2\pi=-3$ THz,  ${\Gamma}\simeq 37$~MHz,   $\delta/2\pi=-11$~THz, and $T=530$~K. The resonant Rabi splitting frequency $\Omega _{0} /2\pi$ is  0.1 THz  for curves (1) and 0.03 THz for curves (2). Initial conditions are: $S_{x,y}(0)=0$ and $S_{z}(0)=-1$. The  vertical line corresponds to the time  $T_{therm}$ of thermalization of coupled atom-light states for  $\Omega _{0} /2\pi=0.1$ THz.}
\end{figure}

The numerical solution of the full set of Eqs.\eqref{eq18} revealing the quasiequilibrium  dynamics of the coupled atom-light system  for 500-bar argon buffer gas is presented in Fig.\ref{fig5}.
We suppose that all atoms initially occupy  the lower dressed-state  level ${\left| 2(N) \right\rangle} $; that is,  $S_{z} (t=0)=-1$.  Figure \ref{fig5}(a) demonstrates the decreasing atomic polarization described by dressed-state coherences, the Bloch vector components  $S_{x}$  and $S_{y}$,  respectively. In particular,  in the steady-state $S_{x,y}$ tend to the values $S^{(st)}_{x,y}$ according to Eqs.(\ref{st_solve}a) and (\ref{st_solve}b). They clearly exhibit the presence of some small (residual) polarization of the atomic medium which  depends on the vital parameter ${\Gamma _{+}/w} $. The dressed-state coherences $S_{x}$ and $S_{y}$  are  relatively  large for nonequilibrium states when  $\Gamma _{+} \gg w$. On the contrary,  we can use Eq.(\ref{eq23}a) and  ignore the dressed-state polarization  at full thermal equilibrium under the secular approximation (\ref{eq22}).

The transient regime of thermalization of coupled atom-light states  for the experiment described in Sec.\ref{secVI}  is presented in Fig.\ref{fig5}(b).
The dependences based on a numerical solution of full Eqs.(\ref{eq18}) are represented by solid curves in Fig.\ref{fig5}(b). Dashed curves correspond to an expression (\ref{eq31}).
The steady-state levels for various resonant Rabi frequencies
$\Omega _{0}^{} /2\pi $ are shown by horizontal (magenta) lines.
These levels are  determined by the magnitude of population
imbalance $S_{z}^{} $  for a detuning $\left|\delta
\right|={k_{B}T/ \hbar } $ that corresponds to a frequency of
about 11 THz.
From Fig.\ref{fig5}(b) it follows that  for sufficiently
long  time scales  such  as  $t \gg {1/(2w+\Gamma _{+})}$, the
dressed-state population imbalance $S_{z} (t)$ reaches its
steady-state value $S_{z}^{(st)} $ (for a given Rabi frequency
$\Omega _{0}^{} /2\pi $) as determined by expression (\ref{st_solve}c).

\section{\label{intensity}INTENSITIES OF SPECTRUM COMPONENTS}
Here we discuss the modification of the intensities of the fluorescence spectrum components under the thermalization of atomic dressed states.
In general the fluorescence triplet consists of a central line of frequency $\omega _{L} $ and  two sidebands centered at $\omega _{L} \pm \Omega _{R} $ (see Fig.\ref{fig3}).  These lines have intensities with weights
\begin{widetext}
\begin{subequations} \label{eqA14}
\begin{eqnarray}
\label{eq14a}I_{11} \equiv I(\omega _{L} +\Omega _{R} )=\sigma _{11} \Gamma _{1\to 2} =\Gamma \cos ^{4} (\theta )\frac{w(\theta )\left[1+S_z^{(eq)}\right]+\Gamma \sin ^{4} \theta }{2w(\theta )+\Gamma _{+} (\theta )},\\
\label{eq14b}I_{22} \equiv I(\omega _{L} -\Omega _{R} )=\sigma _{22} \Gamma _{2\to 1} =\Gamma \sin ^{4} (\theta )\frac{w(\theta ){\rm \; }\left[1-S_z^{(eq)}\right]+\Gamma \cos ^{4} \theta }{2w(\theta )+\Gamma _{+} (\theta )}, \\
\label{eq14c}I_{0} \equiv I(\omega _{L} )=\Gamma \sin ^{2} (\theta)\cos ^{2} (\theta).
\end{eqnarray}
\end{subequations}
\end{widetext}

Following (\ref{eqA14}c) the  $I_{0}$ intensity component  weight  is still unchanged  during the thermalization process because it does not depend on temperature.   In other words thermalization of coupled atom-light states  reduces to redistribution of intensities between the $I_{11}$ and the $I_{22}$ components.

To be more specific, we examine Eq.(\ref{eqA14}) in the perturbative limit (\ref{eq15}) for negative   detuning $\delta <0$.  In this case we can assume that $\sin ^{2} (\theta )\approx {\Omega _{0}^{2} }/{4\delta _{}^{2} } $, $\cos ^{2} (\theta )\approx 1$ [see (\ref{eq13})].   At the same time in this limit we have ${\left| 1(N) \right\rangle} \sim {\left| b,N \right\rangle} $ and ${\left| 2(N-1) \right\rangle} \sim {\left| a,N \right\rangle} $ for dressed  levels of one manifold. Physically this means that $I_{11} $ approximately corresponds  to the transition  from the upper to the lower state (see Fig.\ref{fig3}).

Let us now analyze Eqs.(\ref{eqA14}) in the limit where thermalization does not occur, that is, when condition (\ref{eq27}) is fulfilled for coupled atom-light states.
From \eqref{eqA14} we get
\begin{subequations} \label{eqA15}
\begin{eqnarray}
I_{11} \approx w(1+S_{z}^{(eq)} )\approx \frac{\gamma \Omega _{0}^{2} e^{{-\hbar \left|\delta \right|/ k_{B} T} } }{\delta _{}^{2} },\\
I_{22} \approx \frac{\Gamma \Omega _{0}^{4} }{16\delta _{}^{4} }, \\
I_{0} = \frac{\Gamma\Omega _{0}^{2} }{4\delta _{}^{2} }.
\end{eqnarray}
\end{subequations}
In expressions  \eqref{eqA15} we arrive at familiar results (see, e.g., \cite{12}) in the limit of near-resonant atom-field interaction when $\hbar \left|\delta \right|\ll k_{B} T$ and $I_{11} \approx 2I_{0} \gamma \ / \Gamma \gg I_{0} $.   However, in the general case, from \eqref{eqA15} we obtain
\begin{equation} \label{eqA16}
\frac{I_{11} }{I_{0} } \approx \frac{4\gamma e^{{-\hbar \left|\delta \right|/k_{B} T} } }{\Gamma } .
\end{equation}
For large detuning $\delta$, one can  have $I_{11} \ll I_{0} $ under the condition  $\hbar \left|\delta \right|\gg k_{B} T\ln \left[4{\gamma \mathord{\left/ {\vphantom {\gamma  \Gamma }} \right. \kern-\nulldelimiterspace} \Gamma } \right]$.  However,  such a condition is not realized for the experimentally accessible range of detuning $\delta $ and  temperatures $T$ -- cf.~\eqref{eq14}.

We can show that in the presence of thermal equilibrium, under the conditions  \eqref{eq25} and \eqref{eq26}  the $I_{11}$ intensity component changes essentially, and from (\ref{eqA14}a) we obtain
\begin{equation} \label{eqA17}
I_{11}^{(therm)} \approx \Gamma e^{{-\hbar \left|\delta \right|/ k_{B}T}}.
\end{equation}
At the same time the other intensity weights, $I_{22} $ and $I_{0}$, still remain unchanged  [see Eqs. (\ref{eqA15}b) and (\ref{eqA15}c)]. From (\ref{eqA15}a) and \eqref{eqA17} it is easy to see that
\begin{equation} \label{eqA18}
I_{11} \ll I_{11}^{(therm)} ,
\end{equation}
which implies a significant  increase in $I_{11} $ under the thermalization process.

Now let us turn our attention to positive-valued large atom-field detuning, that is, to $\delta >0$ . Without thermalization ($2w\ll\Gamma $), from (\ref{eqA14}a) and (\ref{eqA14}b) we get  [cf. (\ref{eqA15}a) and (\ref{eqA15}b)]:
\begin{subequations} \label{eqA19}
\begin{eqnarray}
I_{11} \approx \frac{\Gamma \Omega _{0}^{4} }{16\delta ^{4} }, \\
I_{22} \approx \frac{\gamma \Omega _{0}^{2} }{\delta ^{2} }.
\end{eqnarray}
\end{subequations}
On the contrary,  Eqs. (\ref{eqA14}a) and (\ref{eqA14}b)  approach
\begin{subequations} \label{eqA20}
\begin{eqnarray}
I_{11}^{(therm)} \approx \frac{\Gamma \Omega _{0}^{4} }{16\delta ^{4} } e^{-\hbar |\delta |/k_{B} T}, \\
I_{22}^{(therm)} \approx \Gamma,
\end{eqnarray}
\end{subequations}
under the fulfillment of conditions \eqref{eq25} and \eqref{eq26}.

Thus, thermalization  of coupled atom-light states for positive detuning $\delta $ is mostly due to  dramatic increase  in the  $I_{22}^{}$ intensity weight component and suppression of the $I_{11}$ component simultaneously.

It is also fruitful to introduce the total intensity weight  of fluorescence (absorption) $I$ as a sum of all intensity weights  represented in \eqref{eqA14}; that is,  $I=I_{11} +I_{22} +I_{0} $. With the help of definitions \eqref{eq11} and \eqref{eq16},  from   \eqref{eqA14} we get
\begin{equation} \label{eqA21}
I=\sigma _{bb} \Gamma =\frac{\Gamma }{2} \left(1-\frac{\Gamma \cos ^{2} (2\theta )}{2w+\Gamma _{+} } \right)+\frac{\Gamma w\cos (2\theta ){\rm \; }S_{z}^{(eq)} }{2w+\Gamma _{+} }  ,
\end{equation}
where $\sigma _{bb} ={\left\langle b,N \right|} \sigma {\left| b,N \right\rangle} $ is  the population of the upper (excited) atomic state.

The thermalization of the atom-light field states is determined by last term in \eqref{eqA21} (cf.~\cite{12}).
\begin{figure}
\includegraphics[scale=0.43]{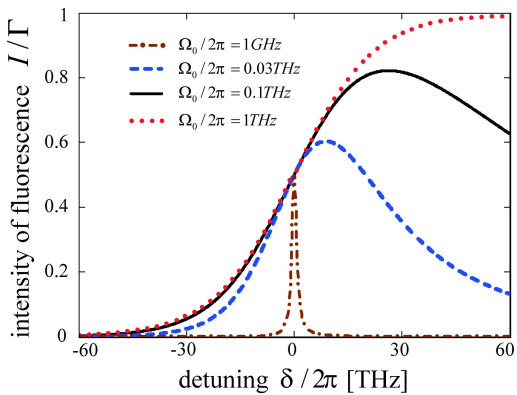}
\caption{\label{fig_int_a} (Color online) The reduced intensity of fluorescence
$I$ versus atom-field detuning $\delta /2\pi $  for 500-bar argon
buffer gas at different  values of the resonant Rabi frequency
$\Omega_0/2\pi$.}
\end{figure}
 In Fig.\ref{fig_int_a} the dependence of the reduced intensity weight $I/\Gamma $ (excited atomic-state population $\sigma _{bb}$) is presented as a function of atom-light field detuning $\delta $.  The asymmetry of the behavior of  $I$ in Fig.\ref{fig_int_a} can be easily understood by introducing  the difference $\Delta I$ between positive (far blue wing)- and negative (far red wing)-valued detuning intensities  for which, from \eqref{eqA21}, we obtain
\begin{equation} \label{eqA22}
\Delta I\equiv I\left(\delta >0\right)-I\left(\delta <0\right)\approx\frac{2\Gamma w\tanh \left({\hbar \left|\delta \right|/ 2k_{B} T} \right)}{2w+\Gamma _{+} } .
\end{equation}

For complete thermalization of the atom-field states [dotted red curve in Fig.\ref{fig_int_a}], under condition (\ref{eq26}) the intensity weight of fluorescence $I$ can be approximated by a Fermi-Dirac distribution function as
\begin{equation} \label{FD_func}
I\simeq\frac{\Gamma}{1+e^{-\hbar\delta / k_{B}T}}.
\end{equation}
In this limit the maximal intensity weight is $I^{(therm)} \approx \Gamma $ (for $\delta >0$) and  $\Delta I$ approaches
\begin{equation} \label{eqA23}
\Delta I^{(therm)} \simeq \Gamma {\rm \; }\tanh \left({\hbar \left|\delta \right|/ 2k_{B} T} \right)\approx \Gamma .
\end{equation}

On the contrary, for small Rabi frequencies (curves describing  nonequilibrium states  in Fig.\ref{fig_int_a}),  from \eqref{eqA22} we have
\begin{equation} \label{eqA24}
\Delta I\approx \frac{\gamma \Omega _{0}^{2} }{\delta ^{2} }\tanh \left({\hbar \left|\delta \right|/ 2k_{B} T} \right)\ll\Gamma ,
\end{equation}
which implies fulfillment of  condition  $\Delta I\ll\Delta I^{(therm)} $  [cf.~\eqref{eqA18}].
\begin{figure}
\includegraphics[scale=0.6]{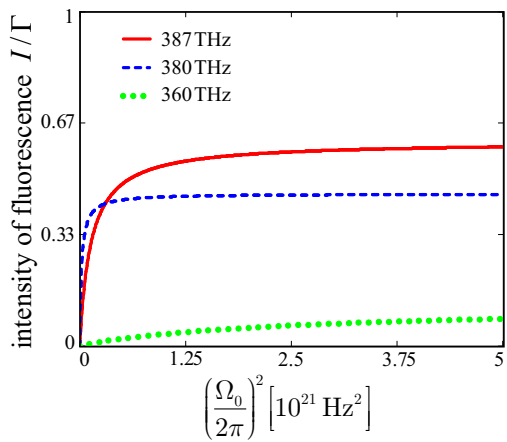}
\caption{\label{fig_int_b} (Color online) The reduced intensity of fluorescence
$I$ versus the square of the resonant Rabi frequency for 400-bar
helium buffer gas at three different frequencies of the laser field.}
\end{figure}

In Fig.\ref{fig_int_b} calculated dependencies of the total
normalized intensity weight $I$ versus the square of the resonant
Rabi frequency (as a measure of the intensity) are presented.
Physically, they reflect the fact that the population
of the atomic ground state ${\left| a \right\rangle} $ decreases, while that of 
the upper level (population $\sigma_{bb}$) increases, with
increases in the incident optical power, that is proportional to the square of resonant  Rabi frequency $\Omega _{0} $ [cf.\eqref{eqA21}].

To end the theoretical part of the paper, let us briefly discuss the boundaries of applicability of our theoretical model for OCs.
It is necessary to accentuate the impact limit ($\Omega _{R}^{} \tau _{coll}^{} \ll 1$) of OCs when the processes of  Rabi oscillations, spontaneous emission, and atomic collisions are mutually uncorrelated. Only in this case is the master equation (\ref{eq2}) used in this paper valid. In our experiment the impact limit is not completely fulfilled (see Sec.\ref{secVI}). In particular, only inequality  $\Omega _{0}^{} \ll\tau _{coll}^{-1} $ takes place in our case. At the same time the experimentally accessible atom-light detuning $\delta $ is of the order of the inverse time of collisions  $\tau _{coll}^{-1} $.

Further, we do not take into account the influence of the Franck-Condon factor  (overlapping  integral) in electric dipole operators, which characterizes overlapping between the (quasi-) molecular wave functions for initial and final states during the collision (see, e.g., \cite{int1}).  Obviously,  this factor becomes more important when the impact limit breaks down  (for detuning $\delta $ such as  $\left|\delta \right|\tau _{coll}^{} \gg 1$)  and a full quantum mechanical approach  to OCs and line profile description become necessary (cf. \cite{36}, \cite{int2}, \cite{int3}).  Nevertheless, as we will see below, the dependences in Figs. \ref{fig_int_a} and \ref{fig_int_b} obtained under the proposed  theoretical approach, are qualitatively in good  agreement with obtained experimental results.

\section{\label{secVI}EXPERIMENTAL PART: THERMAL EQUILIBRIUM FOR COUPLED ATOM-LIGHT STATES}
We now proceed to experimental work, in which we have investigated 
the rubidium atoms under an extraordinarily high buffer gas pressure
driven by tunable laser radiation. The aim of our work presently
is to reach thermal equilibrium of coupled atom-light states.
Clearly, thermal equilibrium is a prerequisite for observation
of a Bose-Einstein-like phase transition for polaritons. The large
collisional broadening in our system interpolates between the usual
atomic gas phase and the solid or liquid phase condition.

\subsection{Experimental setup}
\begin{figure}
\includegraphics[scale=1]{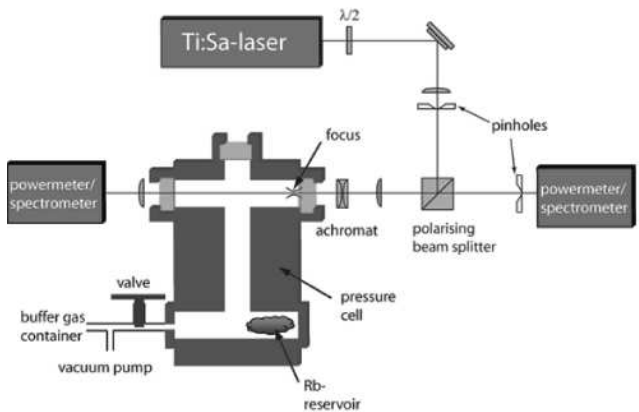}
\caption{\label{fig6} Experimental setup.}
\end{figure}

A scheme of our experimental setup used to investigate atom-light interaction at ultrahigh buffer gas pressures and intense laser radiation is shown in Fig.\ref{fig6}. A stainless-steel high-pressure cell filled with atomic rubidium is used, which is connected by a valve to a buffer gas reservoir. Alternatively helium or argon buffer gas at pressures up to 230~bar is filled into the cell at room temperature. By heating the sealed cell up to  530~K, a buffer gas pressure of 500 bar for argon and a rubidium density of $10^{16}$~cm$^{-3}$ are reached. Tunable laser radiation derived from a titanium-sapphire laser tuned to the rubidium D lines is focused on the cell to a waist size of $3 \mu$m. The focus is placed directly behind the window to suppress propagation effects. To detect fluorescence light selectively from the focal region, where high laser intensities, up to $10^{9}$~mW$/$cm$^{2}$,   for an optical power of 300~mW  are reached, both the incident beam and the outgoing fluorescence are spatially filtered with pinholes in a confocal geometry.
\subsection{Excited-state lifetime measurements}
\begin{figure}
\includegraphics[scale=1.2]{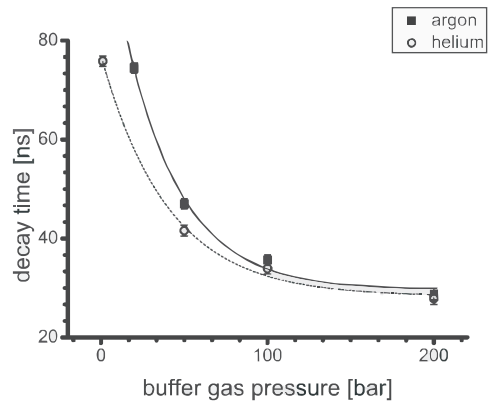}
\caption{\label{fig7}Observed fluorescence decay time of the 5P rubidium excited state versus buffer gas pressure. Data recorded for argon (squares) and helium (circles) gas together with exponential decay curves fitted to the data for helium (dashed line) and for argon (solid line).}
\end{figure}

In initial experiments, we have measured the lifetime of the rubidium 5P excited state in the presence of high buffer gas pressures. The lifetime of excited states is a major limitation of the coherence time in semiconductor polariton BEC experiments, which is here typically close to 1 ps~\cite{1}. For the excited-state lifetime measurement in our rubidium high-pressure buffer gas system, we have chopped the exiting Ti:sapphire laser beam with an acousto-optical modulator (not shown in Fig.\ref{fig6}) and measured the fluorescence in the  backward direction with a photomultiplier. Measurements of the 5P-5S fluorescence decay time for various buffer gas pressures of argon and helium are shown in Fig.\ref{fig7}. For low pressures the decay time is significantly longer than the natural 5P lifetime of 27 ns~\cite{29}. We ascribe the slow decay of the fluorescence signal to energy pooling~\cite{30}, where in excited-atom/excited-atom collisions two Rb atoms pool their internal energy to produce one ground-state atom and one in a higher excited state. A highly excited atom with a long lifetime can subsequently decay into the 5P state. Therefore energy pooling yields a reoccupation process of the 5P state that increases the observed 5P-5S fluorescence decay time. For higher buffer gas pressures the observed decay time is reduced  and tends to approach the expected 5P natural lifetime. We attribute this to a decline in the energy pooling process that involves highly excited rubidium states at high buffer gas pressures, which reduces the efficiency of the repopulation process at high pressure values.
This interpretation of the measured decay times is supported by the observed strong blue fluorescence near 420 nm, which gives clear evidence of the presence of energy pooling, at low buffer gas pressures. Further, with increasing pressure the intensity of the blue fluorescence decreases considerably.

To guide the eye, exponential decay curves have been fitted to the lifetime measurement data. The graphs approach a decay time of 28,3~ns $\pm$ 1,9~ns for helium and 29,6~ns $\pm$~2,2~ns for argon at high pressure values. These values would coincide with the natural lifetime within our experimental accuracy.

No evidence of a decline in the 5P lifetime, possibly to quenching, was observed within our present measurement accuracy up to the quoted pressure value of 200 bar, which was the maximum available buffer gas pressure in these data sets. In future, we plan to extend these lifetime measurements toward higher buffer gas pressures.

\subsection{Measurements indicating thermal quasiequilibrium of atom-light states}

\begin{figure}
\includegraphics[scale=0.7]{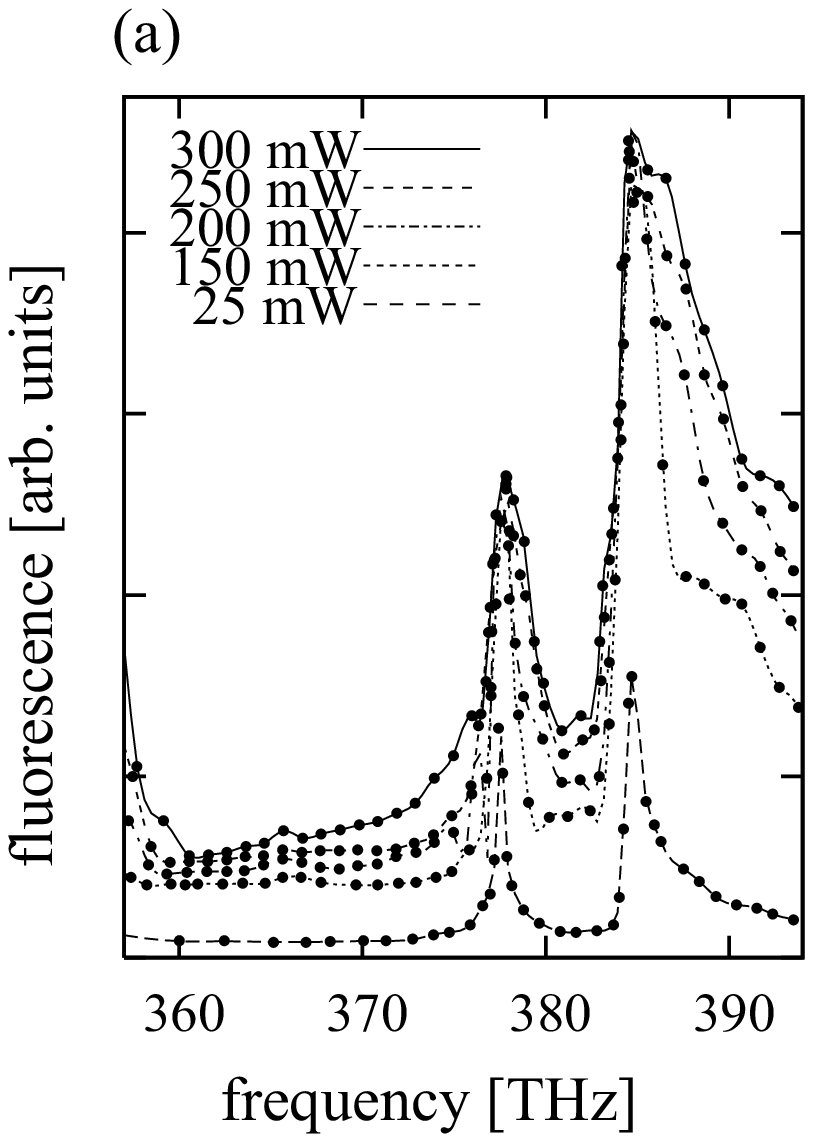}\\
\includegraphics[scale=0.6]{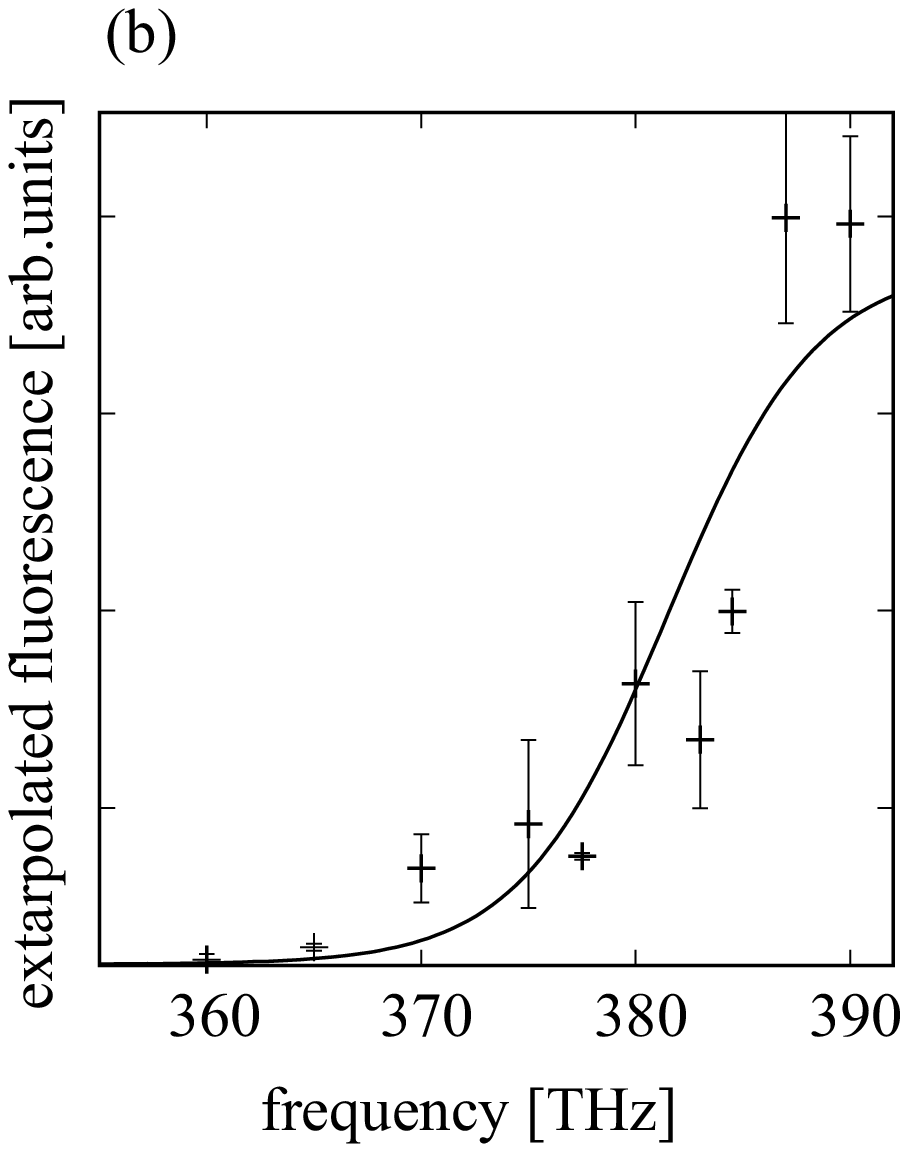}\\
\caption{\label{fig8} (a) Rubidium fluorescence signal and (b) 
results of extrapolation of measurement data to infinite laser
intensity as a function of the incident frequency: both for helium
buffer gas. Error bars were obtained from the
extrapolation fits. Data were fitted with a Fermi-Dirac
distribution, assuming thermal equilibrium of the two-level
dressed-state system [cf. (\ref{FD_func})].}
\end{figure}

In subsequent measurements we have recorded spectra of the rubidium D lines at ultra-high buffer gas pressures. Typical fluorescence spectra recorded at 530~K and 400~bar helium buffer gas pressure for variable optical powers are shown in Fig.\ref{fig8}(a).
Since the fluorescence is measured in the experiment, the population of dressed  states  and population of  excited atomic level $\sigma_{bb}$ [see \eqref{eqA21}] are of interest as well.

At a moderate optical power (P = 25 mW) the usual pressure-broadened rubidium $D_1$ and $D_2$ lines are visible. The resonant Rabi frequency  $\Omega _{0}^{} /2\pi $ is about 0.03 THz  in this case and we deal with state (\ref{eq3}), which is placed far from the region of the resonant atom-field interaction (see Fig.\ref{fig2}). For atom-field  detuning ${\delta/2\pi}=-11$~THz  and for collisional broadening ${\gamma _{Ar}/2\pi} \simeq 3.6$~THz the minimal thermalization time occurs for 500~bar argon buffer gas pressure and $T_{therm} \approx 37.4$~ns in this case, which is  of the order of spontaneous emission lifetime $\tau _{spont} =27$~ns. Thus, we have not yet reached thermal equilibrium  for such a state, represented  by dashed (blue) curves in Figs. \ref{fig4} and  \ref{fig5}.

In particular, at the experimentally achieved maximal power $P\simeq 300$~mW,  the resonant Rabi frequency is   $\Omega _{0}^{} /2\pi \simeq 0.1$~THz and the time of thermalization $T_{therm}$ approaches $10.8$~ns (for 400-bar helium buffer gas pressure) and $3.37$~ns (for 500-bar argon buffer gas pressure). The latter value is  essentially shorter  than the spontaneous emission lifetime $\tau _{spont}$.  In this sense  we can  speak about achieving   equilibrium for coupled atom-light states [state (2) in Fig.\ref{fig2}]  in the experiment (see Fig.\ref{fig5}).

\begin{figure}
\includegraphics[scale=0.6]{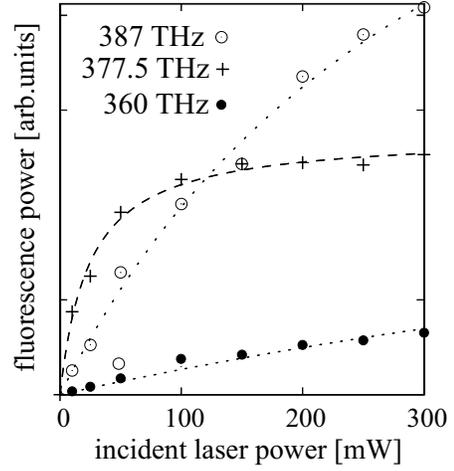}
\caption{\label{fig9}Fluorescence signal at 400-bar helium buffer gas as a function of input laser  power  for three different values of laser frequency.}
\end{figure}

Further diminishment of the thermalization time is possible by increasing the incident optical power.  Notably, full thermalization of the coupled atom-light state is achieved for infinite laser power (cf. Fig.\ref{fig4}) or full line saturation (see Fig.\ref{fig9}).

For the purpose of comparison the expected fluorescence signal at full saturation versus frequency is shown in Fig.\ref{fig8}(b). It was derived at each frequency point by extrapolating the observed fluorescence to infinite laser intensity. Interestingly, this spectrum of the extrapolated fluorescence can be fitted well by a Fermi-Dirac function $I(\delta)$ (solid line) [see (\ref{FD_func}]. We interpret these measurements as evidence for approach of dressed-states thermal equilibrium at high drive laser powers (cf. Fig.\ref{fig_int_a}). Strictly speaking, only in this case does the Franck-Condon factor (overlapping of quasimolecular wave functions) not matter.

Figure  \ref{fig9} shows the experimentally observed fluorescence versus optical power for three different laser frequencies for the helium buffer gas data.
Near a line center (crosses) the curve has already saturated at a relatively low power level to an intermediate fluorescence level. For significant detuning, the optical power at which saturation is achieved is higher, as visible for both blue (open circles) and red (filled dots) detunings.

Interestingly, the saturation level of fluorescence is clearly different in both cases, with the blue (red)  detuning leading to the largest (smallest) values. At full saturation, where spontaneous processes are negligible compared to stimulated processes, we expect the dressed-state populations to be in full  thermal equilibrium (cf.~Figs. \ref{fig4} and \ref{fig_int_b}).

During the course of the experiments, we found that the results of the spectral measurements were critically dependent on the purity of the buffer gas used and presumably, also, the decomposition of residual impurities.
The measurements shown in Fig.\ref{fig8} were carried out using the buffer gas argon/helium 5.0 (supplier: Air Liquide). In contrast, we did not observe significant saturation of the fluorescence at high laser powers for buffer gas from a different supplier (argon/helium 5.0; supplier, Praxair). We attribute this to different residual impurity composition. In a pure inert buffer gas system this perturbation apparently does not occur at a significant rate, which reveals the remarkable elasticity of alkali gas atoms in collisions with inert gas atoms. To experimentally reach full thermal equilibrium also at very large laser detuning it is necessary to enlarge the coupling in the atom-light system. This can be achieved by use of a higher laser power or a cavity-based system.

\subsection{Toward thermalization of atom-light states with a continuum of modes}
Up to now thermal equilibrium of a coupled atom-light system has been
investigated for the case of a two-level dressed-state system. The
observation of a Bose-Einstein-like phase transition from a
thermal state to an ordered polariton state, however, requires a
continuum of modes (see, e.g.,~\cite{31}--\cite{25}), as, for example, provided by the
transverse modes of an optical cavity. Alternatively, a
suitable optical waveguide could be used. To allow 
accumulation of polariton modes at the lowest available energy, it
is clear that a process must exist that allows a continuous
change of photon wavelengths toward lower energies, at least in
the region around the lowest available photon energy of the
optical resonator.

\begin{figure}
\includegraphics[scale=0.3]{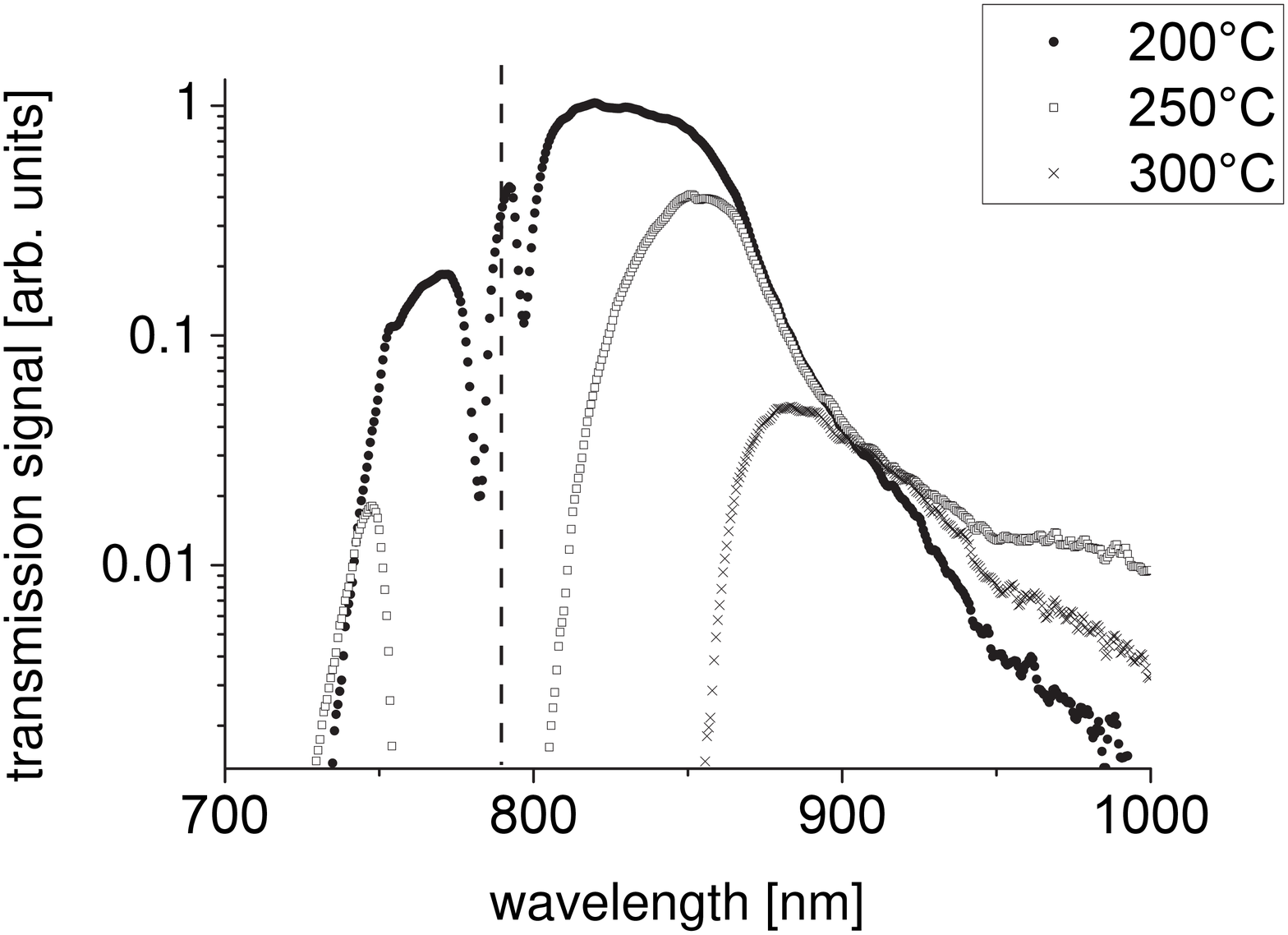}
\caption{\label{fig10}Relative optical intensity measured in the
forward direction after the cell versus frequency for different
cell temperatures. The value of the incident laser wavelength (790
nm) is indicated by the dashed verticalline.}
\end{figure}

In our rubidium high-pressure buffer gas system, we have tested for the presence of such frequency shifting processes. For this experiment, we spectrally analyzed the radiation detected at the cell output in the forward direction behind a second window of the cell (see Fig.\ref{fig6}). Note that in this forward-direction measurement, because of the comparatively long optical path through the cell (5 cm), the confocal parameter used ($\simeq 70\mu m$) is much shorter than the interaction length, so that the beam diameter  increases considerably at the end of the cell and the optical intensity there is comparatively low. Due to the long interaction length, we also expect propagation effects to play a role. Typical spectra, measured with a grating spectrometer, are shown in Fig.\ref{fig10} for various cell temperatures. The incident laser wavelength here was 790 nm, that is, between the rubidium $D_1$ and the rubidium $D_2$ lines. At moderate temperatures ($200^{\circ } C$) the observed fluorescence spectrum covers the spectral range from 750 nm to 950 nm, with clear dips near the rubidium D lines, which we attribute to the greater absorption there.
At higher temperatures and therefore higher Rb densities, the cell
becomes optically thick over a large spectral range near the
rubidium resonances, and in this wavelength range little transmitted
optical power is observed. The spectra show a broad fluorescence
peak with a maximum wavelength exceeding 900 nm. The fluorescence
signal observed at the cell output has maximum spectral intensity
on the red side of the rubidium spectrum; that is, Stokes scattering
is clearly stronger than anti-Stokes processes. We interpret the
results of these measurements as evidence that processes exist
which could support thermalization of photonic modes also in the
case of a continuous mode spectrum, as required for 
experimental observation of a BEC-like phase transition of
polaritons. Clearly, the efficiency of these wavelength shifting
processes, and also thermalization of polariton modes, needs to be
studied in detail in future work. For the rubidium buffer gas
system, a reasonable choice of wavelength for the lowest-cavity 
mode seems to be between 800 nm and 900 nm, where optical modes
are populated by Stokes scattering. The presence of the phase
transition would change the broad fluorescence signal in the forward
direction to a sharp peak located at the frequency of the lowest
mode.

\section{Conclusions and outlook}\label{secVII}
In this paper have we studied the thermalization of coupled atom-light
states under the influence of OCs. We found that
thermal equilibrium is possible by controlling the resonant Rabi
frequency and the atom-light detuning. We have shown that a
nonvanishing macroscopic polarization of the atomic medium occurs
when the problem is described without the secular approximation,
which is typically used in the problem under discussion.
Experimentally we find evidence for a thermalization of the
dressed atom-light states in an ultrahigh-pressure buffer gas 
environment. The observed intensity-dependent asymmetry of the
spectra is interpreted as partial thermal equilibrium of dressed
atom-light states, when the driving-field detuning is chosen as
$\left|\delta \right|\simeq {k_{B}T/\hbar }$. The thermalization
process results in significant energy redistribution within the
two sideband intensity components. We have observed Stokes
scattering and characterized the lifetime of excitations in the
presence of the buffer gas.

In the future, it will be important to add a spatial confinement
to allow dispersion of the coupled atom-light eigenstates
with a low-frequency cutoff, suitable for BEC. 
This can be implemented using either a resonator or
a waveguide structure. This also yields an enhancement of the
field amplitude, which can allow full thermalization of
coupled atom-light states in the high-pressure buffer gas system.
When the rubidium density is increased, a strong coupling limit
should be achievable in the buffer gas system, making polaritons
relevant atom-light excitations. On the theoretical
side it will be important to extend the treatment of coupled
atom-light state (polariton) thermalization in the presence of OCs, 
as done in the present work.

We conclude that the ultrahigh-pressure buffer gas  approach is a
promising candidate for possible realization of a BEC-like
phase transition of polaritons in an atomic physics system.

\section*{ACKNOWLEDGMENTS}
This work was partially supported by Russian Foundation for Basic
Research Grant No. 09-02-91350 and by the Deutsche
Forschungsgemeinschaft within FOR 557 and cooperation project
436 RUS 113/996/0-1. We are grateful to the referees for their comments and the referral the valuable Ref. \cite{36}. A.P. Alodjants acknowledges financial support from Russian Federal Agency of Education under the program ``Scientific and scientific-pedagogical potential of Russia for innovation''.

\end{document}